# The most stable isomer of $H_2C_4$-$(OCS)_2$ van der Waals complex: Theory and experiment agree on a structure with $C_2$ symmetry

**A.J. Barclay,[1] A. Pietropolli Charmet,[2] N. Moazzen-Ahmadi[1,a]**

[1] Department of Physics and Astronomy, University of Calgary, 2500 University Drive North West, Calgary, Alberta T2N 1N4, Canada

[2] Dipartimento di Scienze Molecolari e Nanosistemi, Università Ca' Foscari Venezia, Via Torino 155, I-30172, Mestre, Venezia, Italy

[a] Corresponding author. Electronic mail: ahmadi@phas.ucalgary.ca

**Abstract**

We report the infrared spectrum of $H_2C_4$-$(OCS)_2$ trimer in the region of the $\nu_1$ fundamental vibration of the OCS monomer. The van der Waals complexes are generated in a supersonic slit-jet apparatus and probed using a rapid-scan tunable diode laser. Both $H_2C_4$-$(OCS)_2$ and $D_2C_4$-$(OCS)_2$ are studied. Analysis of their spectra establishes that the trimer has $C_2$ point group symmetry. Theoretical calculations performed to find stationary points on the potential energy surface confirm that the observed structure is the most stable form. The experimental rotational parameters are in very good agreement with those computed using double hybrid functionals.

# 1. INTRODUCTION

Mixed dimers and trimers formed from OCS and the first member of the polyyne series, acetylene, have been studied extensively by high resolution spectroscopy. Two isomers of the mixed dimer $H_2C_2$-OCS were first studied by Peebles and Kuczkowski using microwave spectroscopy [1,2]. One of them [1] was observed to be planar and near-parallel analogous to the observed $H_2C_2$-$N_2O$ and $H_2C_2$-$CO_2$ dimers. The other one [2] was also planar but T-shaped similar to acetylene dimer. The T-shaped isomer has $C_{2v}$ symmetry, with OCS forming the stem of the T, and the S atom in the inner position. Infrared spectra of both dimers were later studied in the OCS $\nu_1$ region [3] and they showed rather different vibrational shifts, a small red shift (-0.3 cm$^{-1}$) for the near-parallel form and a larger red-shift (-5.7 cm$^{-1}$) for the T-shaped form.

The first identification of the mixed trimer $H_2C_2$-$(OCS)_2$ was also made by Peebles and Kuczkowski [4]. This trimer (isomer b) was found to possess a twisted barrel shaped structure similar to $(OCS)_3$ [5] and $(N_2O)_3$ [6]. It has a polar OCS dimer unit, hence its rotational spectrum was studied by means of microwave spectroscopy. However, the pure rotational spectrum of the ground state isomer of $H_2C_2$-$(OCS)_2$ with a non-polar OCS dimer unit could not be detected due to a very small permanent dipole moment which could arise only from induced effects. This lowest energy isomer (isomer a) was later observed in the infrared region [7]. It has a $C_2$ symmetry axis perpendicular to and passing through the center of mass of $C_2H_2$, in contrast to isomer b which has no symmetry. Two fundamental bands for isomer a were observed in the OCS $\nu_1$ region. One of the bands is a relatively strong $c$-type band associated with the out-of-phase vibration of the OCS monomers. The other band, with $b$-type selection rules, arises from their in-phase vibration and would have zero intensity in the planar limit. Observation of a $b$-type band thus establishes the nonplanarity of the OCS dimer unit within the trimer.

A related trimer is $(C_2H_2)_2$-OCS. Two distinct isomers of this trimer have been observed. First is a barrel-shaped isomer whose precise structure is still unclear [8]. This uncertainty is probably due to the presence of large amplitude intermolecular motions. The second isomer is a planar form with the $C_2H_2$ monomers in a nearly T-shaped orientation, like $C_2H_2$ dimer, and OCS approximately parallel to the 'stem' of the T [9]. The energy ordering of the two isomers has not been well established.

Much less is known on complexes containing OCS and the second member of the polyyne series, diacetylene. The only study is that of the mixed dimer of $H_2C_4$-OCS in the infrared region reported recently by our group [10].  Like $H_2C_2$-OCS, $H_2C_4$-OCS was found to have a planar structure with nearly parallel monomer units, and with the help of theoretical calculations, it was established that the observed structure is the lowest energy form on the potential energy surface.

In the present letter, we report observation of a fundamental band for $H_2C_4$-$(OCS)_2$ in the region of the $v_1$ fundamental vibrations of the OCS monomer. This is a hybrid band with a very strong $c$-type component accompanied by a much weaker $a$-type. The trimer observed here has $C_2$ point group symmetry with the $C_2$ symmetry axis perpendicular to and passing through the center of mass of HCCCCH. We also carried out various levels of theoretical calculations in support of our experimental findings. The theoretical calculations confirm that the observed trimer is the most stable form and has an OCS dimer unit which is planar and non-polar. This is in contrast to $H_2C_2$-$(OCS)_2$ where the OCS dimer has a substantial dihedral angle giving rise to a $b$-type band due to the in-phase vibration of the OCS monomers. In addition to the normal isotopologue, we also study a band for $D_2C_4$-$(OCS)_2$. The results were found to be consistent with those of the normal species. The experimental rotational parameters are in very good agreement with those computed using double hybrid functionals.

## 2. COMPUTATIONAL DETAILS

The experimental investigation was supported by theoretical calculations to characterize the different minima on the potential energy surface (PES) of $H_2C_4$-$(OCS)_2$, and for determining their relative binding energies. Employing different basis sets, several computational approaches were used, from density functional theory (DFT) to wavefunction-based methods, to the coupled cluster (CC) level of theory. When modelling molecular complexes, the dispersion forces must be included in the DFT analysis [11,12]; in the present work all the DFT calculations were carried out including the DFT-D3BJ dispersion corrections proposed by Grimme [13, 14] and the maug-cc-pVTZ basis set [15].

The stationary points on the PES were first characterized by using the B3LYP functional [16, 17], and then further refined using the B2PLYP functional [18]. Subsequent calculations of the harmonic force fields confirmed that all the structures correspond to true minima on the PES. To minimize the basis set superposition error (BSSE) all the DFT calculations employed the counterpoise correction (CP) proposed by Boys and Bernardi [19]. For the most stable isomer, the vibrational corrections to the corresponding equilibrium rotational constants were obtained (within the VPT2 framework) from the anharmonic force field data computed using the B3LYP functional, because of its good performance in modelling this anharmonic part of the potential [20, 21]. At this level of theory, when including in the VPT2 treatment all the intermolecular motions, no negative frequencies for the fundamentals were found, and no abnormally large values of the cubic force constants were detected, thus suggesting that the bias introduced should not be too relevant. Anyway, to overcome this source of errors, in the present work the vibrational corrections were computed using a reduced-dimensionality scheme, and all the modes falling below 60 cm$^{-1}$ were excluded from the VPT2 calculations.

The values of the binding energies were further refined by a composite scheme relying on a series of single-point calculations carried out by using both MP2 [22] and CCSD(T) [23] levels of theory. These in turn rely on different extrapolation schemes to the complete basis set (CBS) limit [see, for example, refs. [24-26], and references therein]. In the present work, the BE estimates were obtained by correcting the MP2/CBS values with the cc-pVTZ and cc-pVQZ basis sets [27-29] with the CCSD(T)/cc-pVTZ energies. Core-valence (CV) corrections were obtained at MP2 level and using the cc-pCVTZ basis set [30].

All the calculations were carried out employing the Gaussian suite of quantum chemical programs [31]; for DFT functionals, the *UltraFine* grid as available in Gaussian16 (corresponding to 99 radial and 590 angular points) was used, because of its good performances (see for example Ref. [32]) for the calculations of anharmonic force field data. We found eight different isomers of $H_2C_4$-$(OCS)_2$ of which four of the lowest energy structures are illustrated in Figure 1.

The most stable form of $H_2C_4$-$(OCS)_2$, with $C_2$ symmetry, is shown in Fig. 1a. It consists of a nonpolar planar OCS dimer unit with a diacetylene on top. The OCS dimer unit has a very small slip angle, similar to the geometry of the lowest energy isomer of the isolated OCS dimer [33,34]. Isomer a is the form observed in the present work. The other calculated isomers in Fig. 1 have not been observed experimentally. Isomer b with a binding energy of -1664 cm$^{-1}$ is located 71 cm$^{-1}$ higher in energy relative to isomer a. The OCS dimer fragment in this case is polar and non-planar. This is in contrast to the planar geometry of the isolated polar OCS dimer [35,36]. Isomer c, at an energy of -1616 cm$^{-1}$, also has $C_2$ symmetry with a non-planar dimer unit. Isomer d with a binding energy of -1560 cm$^{-1}$ consists of a polar and non-planar OCS dimer unit similar to isomer (b) but with a larger slip angle. The calculated equilibrium rotational constants computed from the optimized geometries together with the binding energies for isomers a to d are listed in Table 1 and the Cartesian coordinates for these structures are given in Table A1, which is available as supplementary data associated with this letter.

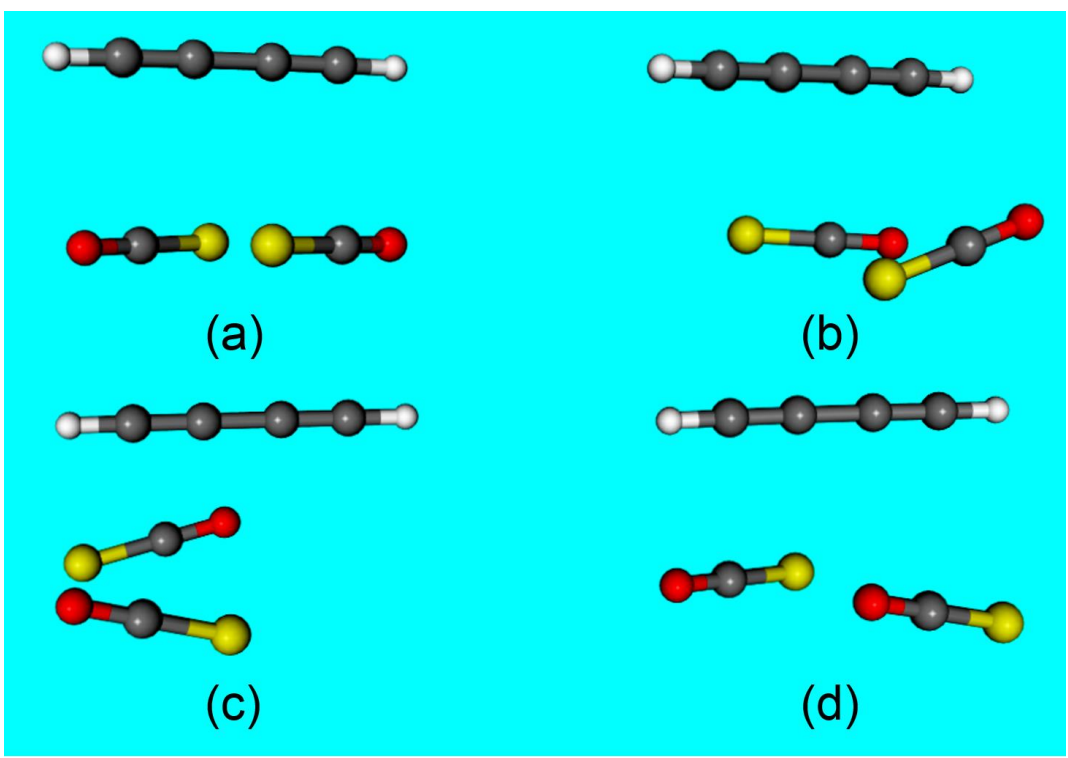

Figure 1: The four lowest energy isomers of $H_2C_4$-(OCS)$_2$ computed using double hybrid functionals. (a) The lowest energy isomer with $C_2$ symmetry and a planar OCS dimer unit studied in this work. (b) Isomer having a polar and non-planar OCS dimer unit. (c) Isomer consisting of a nonpolar and non-planar OCS dimer unit with $C_2$ symmetry. (d) Isomer having a polar and non-planar OCS dimer unit similar to isomer (b) but with a larger slip angle.

Table 1. Theoretical molecular parameters and binding energies (BE) for the four lowest energy isomers of $H_2C_4$-(OCS)$_2$.[a]

|  | $H_2C_4$-(OCS)$_2$ | | | |
|---|---|---|---|---|
|  | Isomer a | Isomer b | Isomer c | Isomer d |
| $A$ / MHz | 866 | 861 | 874 | 972 |
| $B$ / MHz | 731 | 724 | 726 | 667 |
| $C$ / MHz | 617 | 571 | 562 | 515 |
| BE[b] / cm$^{-1}$ | -1735 (-1408) | -1664 (-1334) | -1616 (-1289) | -1560 (-1237) |
| BE[c] / cm$^{-1}$ | -1966 (-1640) | -1899 (-1570) | -1847 (-1518) | -1805 (-1483) |

[a] Equilibrium rotational constants computed from optimized geometries (counterpoise corrected) obtained with B2PLYP-D3BJ functional in conjunction with the maug-cc-pVTZ basis set.

[b] Computed at B2PLYP-D3BJ/maug-cc-pVTZ level of theory (in parentheses the values corrected by ZPV).

[c] Computed using the composite scheme described in the text (in parentheses the values corrected by ZPV).

## 3. RESULTS AND ANALYSIS

The spectra were obtained using a pulsed supersonic slit-jet expansion of a dilute mixture of $C_4H_2$ (0.3%) and OCS (0.1%) in helium carrier gas, with a backing pressure of about 8 atmospheres. All spectra were recorded by direct absorption using a rapid-scan tunable diode laser spectrometer, as described previously [37]. The diacetylene was synthesized by the procedure described in Ref. [38]. DCCCCD was obtained by mixing HCCCCH with a 1 molar solution of NaOD in $D_2O$ as described by Etoh et al. [39]. The samples of diacetylene and diacetylene-$d_2$ thus prepared were purified by repeated distillation under vacuum and then stored at $LN_2$ temperature. The purity of the sample was checked using low resolution infrared spectroscopy. Wavenumber calibration was carried out by simultaneously recording signals from a fixed etalon and a reference gas cell containing OCS. Spectral assignment and simulation were made using PGOPHER software [40].

The most stable isomer (isomer a) with $C_2$ symmetry is depicted with its inertial axes in Fig. 2. Here, the $b$ inertial axis is perpendicular to the diacetylene monomer axis and passes through its center of mass. The angle between the $c$-axis and the equivalent OCS monomers is 23°. Because the $b$-axis is also perpendicular to the OCS dimer plane, we expect only one intramolecular fundamental in the region of the OCS monomer $\nu_1$ fundamental. This band would have a hybrid $a/c$-type structure with a predominating $c$-type component. Due to nuclear spin statistics and the trimer's two-fold symmetry axis, we expect (and observe) a characteristic intensity alternation in the spectrum. In the case of the normal $H_2C_4$-$(OCS)_2$ isotopologue, levels with $K_aK_c$ = ee and oo in the ground vibrational state have a weight of one while other levels have a weight of 3. In the case of $D_2C_4$-$(OCS)_2$, the weights are 6 for ee and oo levels and 3 for eo and oe levels.

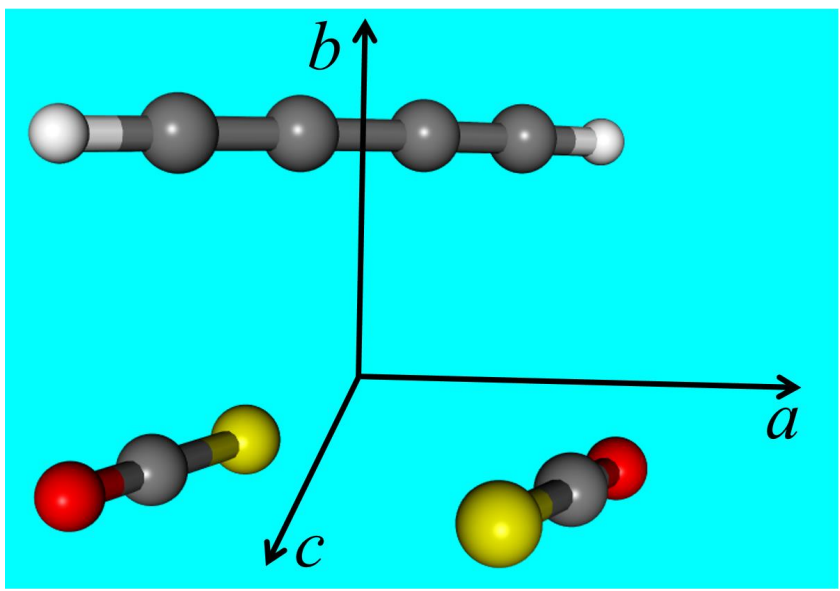

Figure 2: Structure of isomer a of H₂C₄-(OCS)₂ trimer with C₂ symmetry and its principal axes system. Note that the *b*-axis is perpendicular to the diacetylene monomer axis and passes through its center of mass. The OCS dimer fragment is planar. The *c*-axis makes an angle of 23° with the equivalent OCS monomers and the plane of OCS dimer fragment is perpendicular to the *b*-axis.

A search to the higher frequency side of the vibrational fundamental of H₂C₄-OCS dimer [10] resulted in observation of a new band which was assigned to H₂C₄-(OCS)₂. This band was located around 2068.93 cm⁻¹, which is 3.83 cm⁻¹ below the band origin of the polar OCS dimer [35] and 9.30 cm⁻¹ above the H₂C₄-OCS fundamental. As predicted by the theoretical calculations, this vibrational fundamental is observed to be an *a/c*-type band and was assigned to isomer a as the out-of-phase vibration of the pair of equivalent OCS monomers. As with the analogous vibration of the isolated nonpolar OCS dimer, the band exhibits a vibrational blue shift relative to the free OCS monomer, but in this case the shift is smaller, evidently due to the presence of the diacetylene. The regions of the *Q*- and *R*-branches of this band are illustrated in the top trace of Fig. 2. A total of 91 transitions were assigned to 89 observed lines, of which 18 transitions were *a*-type. The majority of transitions were fully resolved with a few which were blended. These line positions could be well fitted in terms of an *a/c*-type band with the spin weights given above. The resulting parameters are given in Table 2.

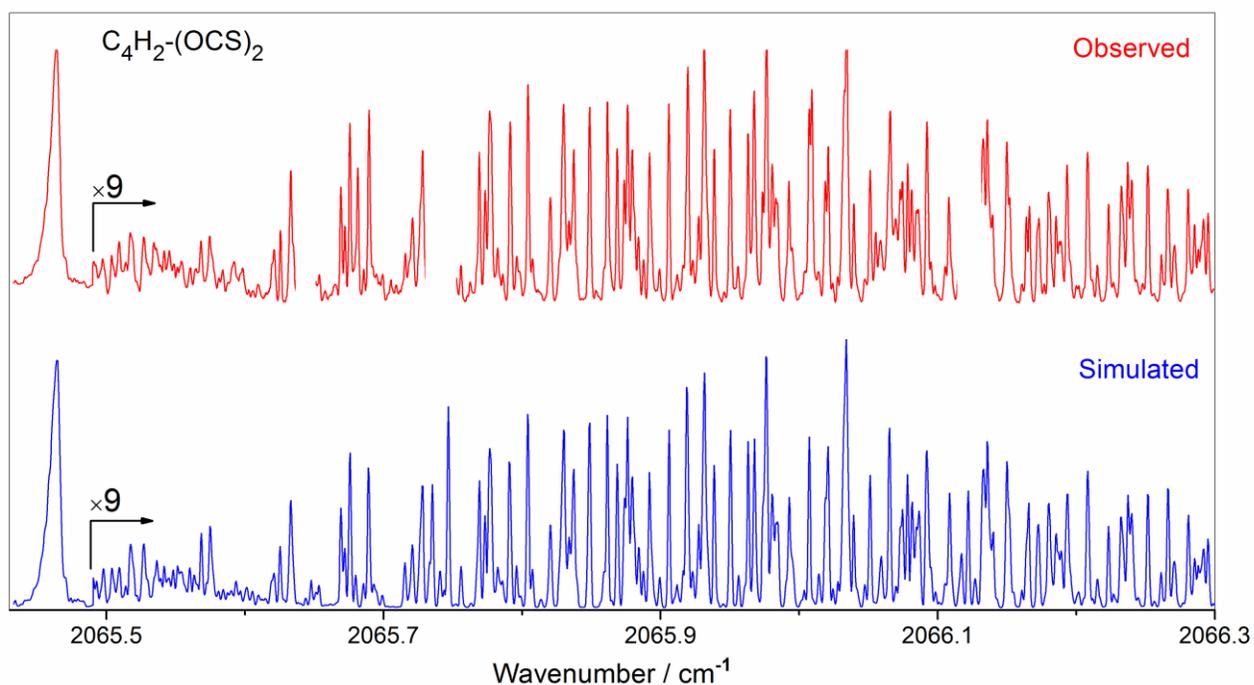

Figure 3: Part of the *a*/*c*-type band of isomer a of the $H_2C_4$-$(OCS)_2$ trimer in the region of OCS $\nu_1$ fundamental. An effective rotational temperature of 3.2 K and an assumed Gaussian line width of 0.0018 cm$^{-1}$ were used for the simulated spectra (bottom trace). The blank regions in the experimental spectrum (top trace) are obscured by absorption due to OCS monomer.

Because of limitations with the DCCCCD sample, the quality of the spectrum for $D_2C_4$-$(OCS)_2$ was not as good as that for the normal isotopologue. However, we still managed to assign 50 frequencies to 79 transitions. These line positions could be fitted, again in terms of an *a*/*c*-type band. The weighted rms of the fit to the seven rotational parameters, listed in Table 2, was 0.00026 cm$^{-1}$ (unweighted rms of 0.00043 cm$^{-1}$). The assignments and observed and calculated line positions for $H_2C_4$-$(OCS)2$ and $D_2C_4$-$(OCS)_2$ are given in Tables A2 and A3 of the supplementary material.

Table 2. Molecular parameters for isomer a of the $H_2C_4$-$(OCS)_2$ trimer in the region of OCS $\nu_1$ fundamental. Uncertainties in parentheses are $1\sigma$ from the least-squares fits in units of the last quoted digit.

| | $H_2C_4$-$(OCS)_2$ | | $D_2C_4$-$(OCS)_2$ | |
|---|---|---|---|---|
| | Experiment | Calculated[a] | Experiment | Calculated[a] |
| $\nu_0$ / cm$^{-1}$ | 2065.4667(1) | | 2065.3180(1) | |
| $A'$ / MHz | 857.17(11) | [854] | 836.31(24) | [832] |
| $B'$ / MHz | 731.118(73) | [727] | 713.69(22) | [710] |
| $C'$ / MHz | 610.475(77) | [610] | 597.55(52) | [598] |
| $D_{JK}'$ / KHz | 1.60(47) | | - | |
| $A''$ / MHz | 856.51(12) | 866 / [854] | 832.97(27) | 842 / [830] |
| $B''$ / MHz | 733.325(84) | 731 / [728] | 715.99(25) | 713 / [710] |
| $C''$ / MHz | 611.622(85) | 617 / [610] | 597.66(53) | 602 / [598] |
| $D_{JK}''$ / KHz | 3.66(52) | | - | |
| $\Delta\nu_0$ [b] / cm$^{-1}$ | +3.266 | | +3.117 | |

[a]Equilibrium rotational constants (at B2PLYP-D3BJ/maug-cc-pVTZ level). The values reported in italic between square parentheses are augmented by vibrational corrections obtained at B3LYP-D3BJ/maug-cc-pVTZ level employing a reduced-dimensionality scheme, see text.

[b]$\Delta\nu_0$ is the vibrational shift of the band origin relative to that of the free OCS molecule.

## 4. DISCUSSION AND CONCLUSION

Comparison of the theoretical equilibrium rotational constants for $H_2C_4$-$(OCS)_2$, columns 2 in Table 1 and columns 3 in Table 2, with the experimental rotational constants, column 2 in Table 2, clearly shows that the observed trimer corresponds to the lowest energy isomer. This is not surprising considering that the jet conditions employed in this work favor the formation of the ground state structure. The comparison with the computed rotational constants augmented by vibrational corrections using the reduced-dimensionality scheme previously reported appears remarkable, but we have to point out that these predictions could be still somewhat affected by some residual errors.

It is also possible to compare the theoretical trimer structure with those of $H_2C_2$-$(OCS)_2$ and the isolated OCS dimer. The OCS dimer subunit within the trimer has R = 3.694 Å, $\theta$ = 84.2°, and dihedral angle = 180°, as compared to values of R = 3.648 Å, $\theta$ = 89.4°, and dihedral angle = 143.9° for $H_2C_2$-$(OCS)_2$ [7] and R = 3.648 Å, $\theta$ = 85.4°, and dihedral angle = 180° for the isolated non-polar $(OCS)_2$ [33]. Here, R is the distance between the centers of mass of the monomers and $\theta$ is the angle between the line connecting the centers of mass and the OCS monomer axes. Also, $\theta$ = 90° means a slip angle of zero and a dihedral angle of 180° indicates that the dimer is planar. We note that $(OCS)_2$ geometries are very similar, with the largest difference being the deviation from planarity of the OCS dimer subunit in the $H_2C_2$-$(OCS)_2$. Another difference is between the orientations of acetylene and diacetylene with respect to the inertial *c*-axis. This angle is nearly zero for $H_2C_2$-$(OCS)_2$, whereas for $H_2C_4$-$(OCS)_2$ it is 49° and tilted toward the hydrogen atoms. As noted in Ref. [7] the nonplanarity of OCS dimer in $H_2C_2$-$(OCS)_2$ is, in part, due to simple steric effects where the larger S atoms are "pushed away" by the approaching acetylene. In the case of $H_2C_4$-$(OCS)_2$, the minimization of steric effects occurs because the longer chain diacetylene can tilt away from the sulfur atoms more easily and at the same time leaving the geometry of the OCS dimer unit unchanged.

The vibrational fundamental of $H_2C_4$-$(OCS)_2$ observed here is analogous to the *c*-type band of $H_2C_2$-$(OCS)_2$ and the allowed infrared band of the nonpolar OCS dimer. Therefore, it is not surprising to see a clear decrease in the vibrational blue shift by about 3 cm$^{-1}$ from +9.81 cm$^{-1}$ for $(OCS)_2$ [**Error! Bookmark not defined.**,41] to +6.73 cm$^{-1}$ for $H_2C_2$-$(OCS)_2$ and to +3.27 cm$^{-1}$ for $H_2C_4$-$(OCS)_2$.

In conclusion, we have observed the ground state isomer of the $H_2C_4$-$(OCS)_2$ trimer by means of high-resolution infrared spectroscopy in the OCS $\nu_1$ fundamental region ($\approx$2060 cm$^{-1}$). The observed rotational structure indicates that the trimer has $C_2$ symmetry, with equivalent OCS monomers and a planar OCS unit, like the isolated nonpolar OCS dimer. Theoretical calculations confirm that the observed structure is the most stable form. The experimental rotational parameters are in very good agreement with those computed using double hybrid functionals. In addition to the normal isotopologue, we also studied a band for $D_2C_4$-$(OCS)_2$ with the results consistent with those of the normal species.

**ACKNOWLEDGEMENTS**

We gratefully acknowledge the financial support of the Natural Sciences and Engineering Research Council of Canada. One of the author (A.C.P.) gratefully acknowledges financial support by University Ca' Foscari Venezia (ADiR funds), and the computational facilities provided by CINECA (grant no.HP10C1RAPR) and by SCSCF ("Sistema per il Calcolo Scientifico di Ca' Foscari", a multiprocessor cluster system owned by Universita' Ca' Foscari Venezia).

**SUPPORTING INFORMATION AVAILABLE**

Geometries for all the isomers are reported in Table A1; detailed assignments and observed and calculated line positions for the infrared bands studied here are given in Tables A2 to A3;.harmonic frequencies and intensities for the most stable isomer are listed in Tables A4 and A5.

Supplementary Material for:

The most stable isomer of H2C4-(OCS)2 van der Waals complex: Theory and experiment agree on a structure with C2 symmetry
A.J. Barclay, A. Pietropolli Charmet,  N. Moazzen-Ahmadi

Table A1: H2C4-(OCS)2 geometries of isomers a-h in Cartesian coordinates (in angstrom)

```
*******************************************************************
H2C4-(OCS)2 (Isomer a)
 C   -1.466682   2.209337   1.195953
 H   -2.288613   2.190731   1.866350
 C   -0.529103   2.222374   0.430712
 C    0.529090   2.222370  -0.430726
 C    1.466669   2.209324  -1.195966
 H    2.288601   2.190712  -1.866363
 S   -1.293930  -0.942669  -1.559809
 C   -1.966085  -0.967490  -0.144915
 O   -2.465162  -0.986689   0.904553
 S    1.293931  -0.942650   1.559812
 C    1.966094  -0.967477   0.144922
 O    2.465177  -0.986681  -0.904543
---------------------------------------------------------------------
H2C4-(OCS)2 (Isomer b)
 C    2.103061   2.617283  -0.663055
 H    2.679218   3.061078  -1.435154
 C    1.445904   2.114095   0.219819
 C    0.701782   1.544566   1.211851
 C    0.036997   1.038343   2.087734
 H   -0.554808   0.589344   2.845380
 S    1.976875  -1.603283  -0.519373
 C    0.697907  -1.897107   0.334396
 O   -0.252998  -2.113480   0.967677
 S   -1.545318   0.293557  -1.570645
 C   -2.312754   0.172272  -0.209682
 O   -2.880340   0.084540   0.800283
---------------------------------------------------------------------
H2C4-(OCS)2 (Isomer c)
 C   -0.779871  -0.320520   2.115890
 H   -0.000403  -0.223133   2.828748
 C   -1.667423  -0.422883   1.298962
 C   -2.664364  -0.533881   0.373803
 C   -3.545896  -0.630312  -0.449437
 H   -4.317713  -0.718458  -1.171613
 S   -0.072218   1.826541  -1.251130
 C    0.682779   2.023371   0.107036
 O    1.242294   2.166758   1.115805
 S    2.306308  -1.287852   0.154104
 C    1.009691  -1.672985  -0.641096
 O    0.053102  -1.958529  -1.232761
```

```
--------------------------------------------------------------------
H2C4-(OCS)2 (Isomer d)
C   -2.320814   1.927328   0.693692
H   -3.233835   1.779399   1.213158
C   -1.274608   2.084506   0.106126
C   -0.090766   2.252462  -0.551322
C    0.962072   2.391904  -1.131960
H    1.884410   2.515647  -1.640617
S   -0.874986  -1.457773  -1.269201
C   -1.876749  -1.409959  -0.064705
O   -2.622025  -1.377568   0.825932
S    2.944024  -0.618900   0.032632
C    1.719428  -0.358347   0.977271
O    0.813704  -0.171886   1.678812
--------------------------------------------------------------------
H2C4-(OCS)2 (Isomer e)
C   -0.084128   1.087685   1.880529
H   -0.695393   0.607088   2.602012
C    0.611962   1.629456   1.051524
C    1.390692   2.238168   0.110709
C    2.079148   2.775792  -0.726687
H    2.680932   3.246413  -1.462484
S   -3.196726   0.020544  -0.007249
C   -1.898593   0.012704  -0.885853
O   -0.939209   0.005286  -1.540368
S    2.386237  -1.443639  -0.335493
C    1.007295  -1.813870   0.305200
O   -0.017788  -2.088236   0.781847
--------------------------------------------------------------------
H2C4-(OCS)2 (Isomer f)
C   -0.267187   2.038175  -1.871017
H   -0.413974   2.021393  -2.921450
C   -0.097331   2.053523  -0.672930
C    0.095955   2.051068   0.677582
C    0.265776   2.031355   1.875609
H    0.412538   2.010689   2.925973
S    2.763873  -0.541812  -0.461813
C    1.707931  -1.057638   0.573020
O    0.923625  -1.439939   1.341315
S   -2.764144  -0.544470   0.460103
C   -1.706882  -1.056701  -0.575169
O   -0.921601  -1.436344  -1.343782
--------------------------------------------------------------------
H2C4-(OCS)2 (Isomer g)
C    1.564030   2.252634  -1.213052
H    2.093654   2.340277  -2.127963
C    0.952749   2.151472  -0.173633
C    0.254468   2.021480   0.991168
C   -0.370733   1.890426   2.019099
H   -0.918765   1.774983   2.919853
S   -2.148773  -1.223926   0.472834
C   -2.079582  -0.064881  -0.582490
O   -2.029442   0.792226  -1.363131
S    1.999447  -1.449685   0.825345
```

```
C      1.374940   -1.293424   -0.602530
O      0.909329   -1.177691   -1.661134
----------------------------------------------------------------------
H2C4-(OCS)2 (Isomer h)
C      2.077487   -2.483574   -1.372771
H      2.162130   -2.980127   -2.306331
C      1.981368   -1.919095   -0.306701
C      1.863792   -1.280189    0.893311
C      1.746172   -0.710890    1.955126
H      1.640690   -0.203703    2.881110
S     -1.582028   -0.797011    0.420961
C     -3.016098   -0.432561   -0.099377
O     -4.077599   -0.160248   -0.487140
S      1.079491    1.718397   -1.323741
C      0.769870    2.051438    0.174609
O      0.540377    2.296607    1.287704
----------------------------------------------------------------------
```

Table A2 Harmonic frequencies and intensities (computed at B2PLYP-D3BJ/maug-cc-pVTZ level of theory) for the most stable isomer of H2C4-(OCS)2.

```
**********************************************************************
```

| Frequency (cm-1) | Intensity (km mol-1) |
|---|---|
| 3478.3 | 0.25 |
| 3477.5 | 170.38 |
| 2236.3 | 0.00 |
| 2088.5 | 892.69 |
| 2070.5 | 0.07 |
| 2064.5 | 8.44 |
| 908.9 | 0.00 |
| 874.9 | 7.69 |
| 874.9 | 0.00 |
| 663.0 | 83.28 |
| 658.2 | 90.27 |
| 657.2 | 4.50 |
| 651.4 | 3.07 |
| 523.6 | 0.00 |
| 521.4 | 3.14 |
| 520.0 | 2.30 |
| 518.3 | 0.60 |
| 517.3 | 4.26 |
| 516.1 | 4.07 |
| 237.9 | 8.04 |
| 236.8 | 7.18 |
| 72.4 | 0.05 |
| 64.2 | 0.08 |
| 61.7 | 0.25 |
| 49.2 | 0.34 |
| 45.4 | 0.34 |
| 42.3 | 0.08 |
| 35.8 | 0.42 |

|       | 33.1 |       | 0.03 |
|       | 28.8 |       | 0.32 |

*********************************************************************

Table A3 Harmonic frequencies and intensities (computed at B2PLYP-
D3BJ/maug-cc-pVTZ level of theory) for the most stable isomer of D2C4-
(OCS)2.

*********************************************************************

| Frequency (cm-1) | Intensity (km mol-1) |
|------------------|----------------------|
| 2691.6 | 0.11 |
| 2685.2 | 96.68 |
| 2115.6 | 0.03 |
| 2088.2 | 898.56 |
| 2070.5 | 0.06 |
| 1929.2 | 4.59 |
| 878.3 | 0.00 |
| 874.9 | 7.74 |
| 874.8 | 0.01 |
| 541.4 | 0.00 |
| 535.1 | 0.43 |
| 525.0 | 41.85 |
| 522.3 | 35.77 |
| 520.9 | 2.05 |
| 519.9 | 9.35 |
| 517.4 | 4.41 |
| 516.0 | 7.42 |
| 482.6 | 0.02 |
| 480.3 | 1.47 |
| 216.9 | 8.92 |
| 216.0 | 7.92 |
| 71.4 | 0.07 |
| 62.0 | 0.09 |
| 61.7 | 0.27 |
| 48.5 | 0.32 |
| 45.0 | 0.33 |
| 41.5 | 0.09 |
| 35.8 | 0.43 |
| 31.5 | 0.02 |
| 28.6 | 0.32 |

*********************************************************************

Table A4: Observed and calculated transitions for H2C4-(OCS)2 in
          the region of the nu1 fundamental band of OCS (units of 1/cm).

```
********************************************************************
  J'  Ka'  Kc'   J"  Ka"  Kc"    Observed    Calculated  Obs-Calc  Weight
********************************************************************
  14   9    5    15  10    5    2064.66914  2064.66931  -0.00017   1.00
  12  11    1    13  12    1    2064.74354  2064.74374  -0.00019   1.00
  12  10    2    13  11    2    2064.75323  2064.75310   0.00012   1.00
  12   8    4    13   9    4    2064.77327  2064.77321   0.00005   1.00
  11   8    4    12   9    4    2064.81970  2064.81953   0.00016   1.00
  10  10    0    11  11    0    2064.84687  2064.84685   0.00001   1.00
  10   9    1    11  10    1    2064.85678  2064.85650   0.00027   1.00
  10   8    2    11   9    2    2064.86638  2064.86641  -0.00003   1.00
   9   8    2    10   9    2    2064.91279  2064.91312  -0.00033   1.00
   9   6    4    10   7    4    2064.93325  2064.93316   0.00009   1.00
   8   8    0     9   9    0    2064.95982  2064.95976   0.00006   1.00
   8   6    2     9   7    2    2064.98053  2064.98053  -0.00000   1.00
  10   2    8    11   2    9    2064.98457  2064.98456   0.00000   1.00
   8   5    4     9   6    4    2064.98963  2064.98949   0.00013   1.00
   8   2    6     9   3    6    2064.99873  2064.99862   0.00010   1.00
   7   7    1     8   8    1    2065.01634  2065.01641  -0.00007   1.00
   7   6    2     8   7    2    2065.02676  2065.02673   0.00002   1.00
   7   4    4     8   5    4    2065.04503  2065.04499   0.00003   1.00
   7   3    4     8   4    4    2065.05796  2065.05783   0.00012   1.00
   6   6    0     7   7    0    2065.07317  2065.07317  -0.00000   1.00
   8   1    8     9   1    9    2065.09090  2065.09095  -0.00005   0.25
   8   0    8     9   0    9    2065.09090  2065.09095  -0.00005   0.25
   6   4    3     7   5    3    2065.09315  2065.09329  -0.00014   1.00
   6   4    2     7   5    2    2065.09627  2065.09628  -0.00001   1.00
   7   2    6     8   2    7    2065.12078  2065.12077   0.00000   1.00
   5   5    1     6   6    1    2065.12998  2065.13001  -0.00003   1.00
   5   4    2     6   5    2    2065.14057  2065.14045   0.00011   1.00
   5   3    3     6   4    3    2065.14874  2065.14887  -0.00013   1.00
   4   4    0     5   5    0    2065.18693  2065.18703  -0.00010   1.00
   4   3    1     5   4    1    2065.19879  2065.19909  -0.00030   1.00
   4   2    2     5   3    2    2065.21001  2065.20977   0.00023   1.00
   3   3    1     4   4    1    2065.24374  2065.24379  -0.00005   1.00
   3   2    2     4   3    2    2065.25257  2065.25267  -0.00010   1.00
   3   1    2     4   2    2    2065.26265  2065.26259   0.00005   1.00
   2   1    1     1   0    1    2065.56823  2065.56836  -0.00013   1.00
   3   2    2     2   1    2    2065.62559  2065.62545   0.00014   1.00
   4   1    3     3   1    2    2065.64798  2065.64770   0.00028   1.00
   4   2    2     3   2    1    2065.65363  2065.65362   0.00000   1.00
   4   2    2     3   1    2    2065.66925  2065.66913   0.00012   1.00
   4   1    3     3   0    3    2065.67210  2065.67208   0.00001   1.00
   4   3    1     3   2    1    2065.67569  2065.67574  -0.00005   1.00
   5   2    4     4   2    3    2065.68575  2065.68574   0.00001   1.00
   4   4    0     3   3    0    2065.68957  2065.68930   0.00026   1.00
   6   0    6     5   0    5    2065.71563  2065.71568  -0.00004   1.00
   5   3    3     4   2    3    2065.72818  2065.72825  -0.00006   1.00
   7   1    7     6   1    6    2065.75588  2065.75592  -0.00003   0.25
   7   1    7     6   1    6    2065.75588  2065.75592  -0.00003   0.25
   6   3    3     5   2    3    2065.76916  2065.76919  -0.00002   1.00
   6   2    4     5   1    4    2065.77335  2065.77329   0.00005   1.00
   7   5    3     6   5    2    2065.78605  2065.78594   0.00011   1.00
   8   0    8     7   0    7    2065.79614  2065.79611   0.00002   1.00
   6   6    0     5   5    0    2065.80427  2065.80413   0.00013   1.00
   8   1    7     7   1    6    2065.80750  2065.80763  -0.00012   1.00
   7   5    3     6   4    3    2065.83728  2065.83735  -0.00007   1.00
   7   6    2     6   5    2    2065.84878  2065.84882  -0.00003   1.00
   7   7    1     6   6    1    2065.86162  2065.86154   0.00007   1.00
   8   4    4     7   3    4    2065.86865  2065.86865   0.00000   1.00
```

```
 8   3   5     7   2   5    2065.87374   2065.87378   -0.00003   1.00
 8   5   3     7   4   3    2065.87619   2065.87621   -0.00001   1.00
 8   5   4     7   4   4    2065.88394   2065.88380    0.00013   1.00
10   1   9     9   1   8    2065.88772   2065.88771    0.00000   1.00
10   2   8     9   2   7    2065.89916   2065.89929   -0.00013   1.00
 8   7   1     7   6   1    2065.90591   2065.90592   -0.00000   1.00
 9   6   4     8   5   4    2065.93869   2065.93870   -0.00000   1.00
 9   7   3     8   6   3    2065.95027   2065.95032   -0.00005   1.00
12   0  12    11   0  11    2065.95581   2065.95611   -0.00030   1.00
 9   8   2     8   7   2    2065.96312   2065.96332   -0.00020   1.00
10   5   5     9   4   5    2065.96751   2065.96754   -0.00003   1.00
10   4   6     9   3   6    2065.97358   2065.97364   -0.00005   1.00
10   3   7     9   2   7    2065.98059   2065.98048    0.00011   1.00
10   7   3     9   6   3    2065.99263   2065.99268   -0.00004   1.00
10   8   2     9   7   2    2066.00719   2066.00729   -0.00009   1.00
10   9   1     9   8   1    2066.02093   2066.02074    0.00019   1.00
11   4   7    10   3   7    2066.02784   2066.02781    0.00002   1.00
11   7   5    10   6   5    2066.03947   2066.03954   -0.00006   1.00
14   1  13    13   1  12    2066.04749   2066.04713    0.00036   1.00
11   8   4    10   7   4    2066.05123   2066.05140   -0.00016   1.00
11  10   2    10   9   2    2066.07830   2066.07826    0.00003   1.00
12   4   8    11   3   8    2066.08115   2066.08102    0.00012   1.00
12   9   3    11   8   3    2066.10811   2066.10826   -0.00015   1.00
12  11   1    11  10   1    2066.13578   2066.13588   -0.00010   1.00
13   8   6    12   7   6    2066.13936   2066.13992   -0.00055   1.00
12  12   0    11  11   0    2066.14993   2066.14971    0.00021   1.00
13   9   5    12   8   5    2066.15184   2066.15208   -0.00023   1.00
16   4  12    15   4  11    2066.16093   2066.16086    0.00007   1.00
13  10   4    12   9   4    2066.16617   2066.16581    0.00036   1.00
14  11   3    13  10   3    2066.22332   2066.22325    0.00007   1.00
16   8   8    15   7   8    2066.26146   2066.26124    0.00021   1.00
16   5  11    15   4  11    2066.28537   2066.28541   -0.00004   1.00
15  13   3    14  12   3    2066.29514   2066.29517   -0.00003   1.00
18   9   9    17   8   9    2066.35823   2066.35823   -0.00000   1.00
*************************************************************************
```

Table A5: Observed and calculated transitions for D2C4-(OCS)2 in
          the region of the nu1 fundamental band of OCS (units of 1/cm).

```
*************************************************************************
 J'  Ka' Kc'   J"  Ka" Kc"   Observed    Calculated  Obs-Calc  Weight
*************************************************************************
 6   6   1     7   7   1    2064.93810   2064.93814   -0.00004   0.25
 6   6   0     7   7   0    2064.93810   2064.93815   -0.00004   0.25
 6   5   2     7   6   2    2064.94741   2064.94721    0.00020   0.25
 6   5   1     7   6   1    2064.94741   2064.94744   -0.00002   0.25
 5   5   0     6   6   0    2064.99298   2064.99256    0.00041   0.25
 5   5   1     6   6   1    2064.99298   2064.99254    0.00043   0.25
 5   4   1     6   5   1    2065.00237   2065.00250   -0.00012   1.00
 5   3   3     6   4   3    2065.00940   2065.00925    0.00014   1.00
 5   1   4     6   2   4    2065.01609   2065.01628   -0.00019   1.00
 5   2   3     6   3   3    2065.01968   2065.01955    0.00012   1.00
 5   3   2     6   3   3    2065.04182   2065.04137    0.00044   1.00
 4   4   0     5   5   0    2065.04738   2065.04724    0.00014   0.25
 4   4   1     5   5   1    2065.04738   2065.04714    0.00024   0.25
 4   2   3     5   3   3    2065.06188   2065.06194   -0.00005   0.25
 4   1   4     5   2   4    2065.06188   2065.06343   -0.00154   0.25
 4   1   3     5   2   3    2065.06895   2065.06858    0.00036   0.25
 4   2   2     5   3   2    2065.06895   2065.06776    0.00118   0.25
 5   0   5     6   0   6    2065.07293   2065.07296   -0.00003   0.25
 5   1   5     6   1   6    2065.07293   2065.07300   -0.00007   0.25
 3   3   0     4   4   0    2065.10202   2065.10224   -0.00022   0.25
```

| | | | | | | | | | |
|---|---|---|---|---|---|---|---|---|---|
| 3 | 3 | 1 | 4 | 4 | 1 | 2065.10202 | 2065.10186 | 0.00016 | 0.25 |
| 3 | 2 | 2 | 4 | 3 | 2 | 2065.10974 | 2065.10995 | -0.00020 | 1.00 |
| 3 | 2 | 1 | 4 | 3 | 1 | 2065.11318 | 2065.11350 | -0.00031 | 1.00 |
| 3 | 0 | 3 | 4 | 1 | 3 | 2065.11652 | 2065.11654 | -0.00002 | 1.00 |
| 3 | 1 | 2 | 4 | 2 | 2 | 2065.11877 | 2065.11905 | -0.00028 | 1.00 |
| 3 | 1 | 3 | 4 | 1 | 4 | 2065.15287 | 2065.15305 | -0.00017 | 0.25 |
| 3 | 0 | 3 | 4 | 0 | 4 | 2065.15287 | 2065.15253 | 0.00033 | 0.25 |
| 2 | 1 | 2 | 3 | 2 | 2 | 2065.16264 | 2065.16299 | -0.00034 | 1.00 |
| 2 | 1 | 1 | 1 | 0 | 1 | 2065.41809 | 2065.41720 | 0.00088 | 1.00 |
| 2 | 2 | 0 | 1 | 1 | 0 | 2065.42356 | 2065.42337 | 0.00019 | 1.00 |
| 2 | 2 | 1 | 1 | 1 | 1 | 2065.42535 | 2065.42557 | -0.00022 | 1.00 |
| 3 | 2 | 2 | 2 | 1 | 2 | 2065.47255 | 2065.47310 | -0.00054 | 1.00 |
| 3 | 3 | 1 | 2 | 2 | 1 | 2065.47998 | 2065.48061 | -0.00062 | 0.25 |
| 3 | 3 | 0 | 2 | 2 | 0 | 2065.47998 | 2065.47932 | 0.00066 | 0.25 |
| 4 | 2 | 3 | 3 | 2 | 2 | 2065.49162 | 2065.49162 | -0.00000 | 1.00 |
| 4 | 2 | 2 | 3 | 1 | 2 | 2065.51551 | 2065.51572 | -0.00020 | 1.00 |
| 4 | 3 | 1 | 3 | 2 | 1 | 2065.52251 | 2065.52231 | 0.00020 | 0.25 |
| 4 | 2 | 3 | 3 | 1 | 3 | 2065.52251 | 2065.52207 | 0.00044 | 0.25 |
| 4 | 3 | 2 | 3 | 2 | 2 | 2065.52663 | 2065.52652 | 0.00011 | 1.00 |
| 4 | 4 | 1 | 3 | 3 | 1 | 2065.53672 | 2065.53644 | 0.00027 | 0.25 |
| 4 | 4 | 0 | 3 | 3 | 0 | 2065.53672 | 2065.53597 | 0.00074 | 0.25 |
| 5 | 2 | 3 | 4 | 1 | 3 | 2065.56614 | 2065.56569 | 0.00045 | 0.25 |
| 5 | 3 | 2 | 4 | 2 | 2 | 2065.56614 | 2065.56666 | -0.00051 | 0.25 |
| 5 | 5 | 0 | 4 | 4 | 0 | 2065.59303 | 2065.59266 | 0.00036 | 0.25 |
| 5 | 5 | 1 | 4 | 4 | 1 | 2065.59303 | 2065.59280 | 0.00023 | 0.25 |
| 7 | 1 | 7 | 6 | 1 | 6 | 2065.60233 | 2065.60233 | 0.00035 | 0.25 |
| 7 | 0 | 7 | 6 | 0 | 6 | 2065.60269 | 2065.60234 | 0.00034 | 0.25 |
| 6 | 2 | 4 | 5 | 1 | 4 | 2065.61790 | 2065.61788 | 0.00002 | 1.00 |
| 6 | 3 | 4 | 5 | 2 | 4 | 2065.62228 | 2065.62260 | -0.00032 | 0.25 |
| 6 | 2 | 5 | 5 | 1 | 5 | 2065.62228 | 2065.62271 | -0.00043 | 0.25 |
| 6 | 4 | 3 | 5 | 3 | 3 | 2065.62669 | 2065.62687 | -0.00017 | 1.00 |
| 6 | 5 | 2 | 5 | 4 | 2 | 2065.63666 | 2065.63679 | -0.00012 | 0.25 |
| 6 | 5 | 1 | 5 | 4 | 1 | 2065.63666 | 2065.63579 | 0.00087 | 0.25 |
| 6 | 6 | 0 | 5 | 5 | 0 | 2065.64976 | 2065.64945 | 0.00031 | 0.25 |
| 6 | 6 | 1 | 5 | 5 | 1 | 2065.64976 | 2065.64949 | 0.00027 | 0.25 |
| 7 | 4 | 3 | 6 | 3 | 3 | 2065.66394 | 2065.66438 | -0.00043 | 0.25 |
| 7 | 3 | 4 | 6 | 2 | 4 | 2065.66394 | 2065.66381 | 0.00013 | 0.25 |
| 7 | 2 | 5 | 6 | 1 | 5 | 2065.67078 | 2065.67041 | 0.00037 | 1.00 |
| 7 | 4 | 4 | 6 | 3 | 4 | 2065.67313 | 2065.67399 | -0.00085 | 0.25 |
| 7 | 1 | 6 | 6 | 0 | 6 | 2065.67313 | 2065.67343 | -0.00029 | 0.25 |
| 7 | 5 | 2 | 6 | 4 | 2 | 2065.67773 | 2065.67762 | 0.00011 | 1.00 |
| 7 | 6 | 2 | 6 | 5 | 2 | 2065.69310 | 2065.69312 | -0.00001 | 0.25 |
| 7 | 6 | 1 | 6 | 5 | 1 | 2065.69310 | 2065.69280 | 0.00030 | 0.25 |
| 7 | 7 | 1 | 6 | 6 | 1 | 2065.70591 | 2065.70643 | -0.00052 | 0.25 |
| 7 | 7 | 0 | 6 | 6 | 0 | 2065.70591 | 2065.70642 | -0.00051 | 0.25 |
| 8 | 4 | 4 | 7 | 3 | 4 | 2065.71058 | 2065.71107 | -0.00048 | 1.00 |
| 8 | 3 | 6 | 7 | 2 | 6 | 2065.72292 | 2065.72302 | -0.00010 | 0.25 |
| 8 | 4 | 5 | 7 | 3 | 5 | 2065.72292 | 2065.72257 | 0.00034 | 0.25 |
| 8 | 5 | 4 | 7 | 4 | 4 | 2065.72659 | 2065.72674 | -0.00015 | 1.00 |
| 9 | 3 | 6 | 8 | 2 | 6 | 2065.76971 | 2065.76952 | 0.00019 | 1.00 |
| 9 | 6 | 3 | 8 | 5 | 3 | 2065.77615 | 2065.77617 | -0.00001 | 1.00 |
| 9 | 7 | 2 | 8 | 6 | 2 | 2065.79301 | 2065.79260 | 0.00040 | 0.25 |
| 9 | 7 | 3 | 8 | 6 | 3 | 2065.79301 | 2065.79316 | -0.00014 | 0.25 |
| 9 | 8 | 2 | 8 | 7 | 2 | 2065.80676 | 2065.80692 | -0.00016 | 0.25 |
| 9 | 8 | 1 | 8 | 7 | 1 | 2065.80676 | 2065.80689 | -0.00013 | 0.25 |
| 10 | 8 | 3 | 9 | 7 | 3 | 2065.84970 | 2065.84999 | -0.00028 | 0.25 |
| 10 | 8 | 2 | 9 | 7 | 2 | 2065.84970 | 2065.84982 | -0.00012 | 0.25 |
| 11 | 6 | 5 | 10 | 5 | 5 | 2065.85872 | 2065.85848 | 0.00024 | 0.25 |
| 11 | 5 | 6 | 10 | 4 | 6 | 2065.85872 | 2065.85852 | 0.00020 | 0.25 |

*******************************************************************************